\documentclass[reprint,aps,amsmath,amssymb]{revtex4}
\usepackage[]{graphicx}
\usepackage{subfigure}
\usepackage{times}
\usepackage{color}

\newcommand{\comment}[1]{}

\newcommand{\be}{\begin{equation}}
\newcommand{\ee}{\end{equation}}
\newcommand{\bea}{\begin{eqnarray}}
\newcommand{\eea}{\end{eqnarray}}

\begin{document}

\title{Exciton formation assisted by longitudinal optical phonons in monolayer transition metal
dichalcogenides}

\author{A. Thilagam}
\email[]{thilaphys@gmail.com}

\affiliation{Information Technology, Engineering and Environment,\\ 
University of South Australia, Australia
 5095.}
%Email: thilaphys@gmail.com}
%\today
%%%%%%%%%%%%%%%%%%%%%%%%%%%%%%%%%%%%%%%%%%%%%%%%%%%%%%%%%%%%% 
\begin{abstract}
We examine a mechanism by which excitons are generated via 
the LO (longitudinal optical) phonon-assisted scattering process 
after optical excitation of  monolayer transition metal dichalcogenides.  The exciton formation time
is computed as a function of the  exciton center-of-mass wavevector,
electron and hole temperatures, and  carrier densities for known values of 
the  Fr\"ohlich coupling constant, LO phonon energy, lattice temperature,  and  the exciton binding energy
 in layered structures. For the monolayer MoS$_2$, we obtain ultrafast exciton formation times on the sub-picosecond time scale at charge densities of 5 $\times$ 10$^{11}$ cm$^{-2}$ and carrier temperatures less than 300 K, in good agreement with recent experimental findings ($\approx$ 0.3 ps).
While excitons are dominantly created at zero  center-of-mass wavevectors at low charge carrier temperatures ($\approx$ 30 K), the exciton formation time is most rapid at  non-zero  wavevectors
at higher temperatures ($\ge $ 120 K)  of  charge carriers. The results
show the inverse square-law dependence of the exciton formation times on the carrier
density, consistent with  a square-law dependence of photoluminescence on the excitation
density. Our results show that excitons are  formed more rapidly
in exemplary monolayer selenide-based dichalcogenides  (MoSe$_2$ and WSe$_2$) than sulphide-based dichalcogenides
(MoS$_2$ and WS$_2$). 

\end{abstract}

\maketitle

%Pacs: 81.05.Hd, 72.10.-d, 72.20.-i, 72.80.Jc
% 71.35 +z, 71.38 +i, 73.40 -c

\section{Introduction}
 
Exciton mediated many-body interactions  give rise  to a  host of  physical effects \cite{qiu2016screening,sahin2016computing,cudazzo2016exciton}
that determine the opto-electronic properties
of low dimensional transition metal dichalcogenides, MX$_2$
(M = Mo, W, Nb and X = S, Se) \cite{scharf2016probing,rama,ugeda2014giant,wang2012electronics,makatom,qiu2013,
splen,komsa2012effects,kormanyos2013monolayer},
with important consequences for  fundamental and applied research.
The confinement of correlated charge carriers or  excitons to a narrow region of space  in low dimensional transition metal dichalcogenides (TMDCs) leads to
unique photoluminescence properties \cite{splendiani2010emerging,plechinger2012low,gao2016localized,ji2013epitaxial,zhu2016strongly,eda2013two,ghatak2011nature,bergh,molinaspin,mai2013many} that  are otherwise absent in  the bulk configurations \cite{saigal2016exciton}.
The availability of state-of-the art exfoliation 
techniques \cite{novoselov2005two,varrla2015large,chen2015nanoimprint,fan2015fast} enable
fabrication of low dimensional transition metal dichalcogenides  that is  useful for applications
 \cite{ou2014ion,li2016charge,perebeinos2015metal,beck2000,tsai2013few,bernardi2013extraordinary, he2012fabrication,wi2014enhancement,bertolazzi2013nonvolatile,ji2013epitaxial,yu2016evaluation,
park2016mos2,eda2013two,radisavljevic2011integrated,lembke2015single,pospischil2014solar,
zhang2014m,yoon2011good}. The excitonic processes that determine the
performance of TMDC-based electronic devices  include
 defect assisted scattering and  trapping by surface  states \cite{shi2013exciton}, decay via exciton-exciton 
annihilation  \cite{shin2014photoluminescence,ye2014exciton,konabe2014effect}, phonon assisted relaxation \cite{thilrelaxjap},  and capture by mid-gap defects through
 Auger processes  \cite{wang2015fast}. Excitonic processes that result in the formation of complex trions \cite{mak,berkelbach2013theory,
thiltrion} and electron-hole recombination with generation of hot carriers \cite{kozawa2014photocarrier} 
are also of importance in device performances.

Dynamical processes incorporating exciton-phonon interactions  underlie
 the opto-electronic properties
of monolayer transition metal dichalcogenides  \cite{jones2013optical}.
The strength of the interactions between coupled charge carriers and phonons
is deduced  from experimental measurements of the  dephasing times \cite{nie2014ultrafast},
exciton linewidths \cite{selig2016excitonic}, 
photoluminescence, \cite{mouri2013tunable} and other parameters such
as the exciton mobility and luminescence rise times.
The exciton formation time is determined by  a complicated interplay of various 
dynamical processes in the picosecond time scale \cite{siantidis2001dynamics}
and is linked to the efficient operation of optoelectronic devices.
To this end, a  comprehensive understanding of how  newly
generated  electron-hole pairs relax energetically  to
form excitons  still remain unclear.
Recently  decay times of $\approx$ 0.3 ps
of the transient absorption signal subsequent to 
the interband excitation of the monolayer  WSe$_2$, MoS$_2$, and MoSe$_2$
 was recorded in time-resolved measurements \cite{ceballos2016exciton}.
The ultrafast decay times were
deduced as the exciton  formation times from electron-hole pairs in monolayer systems.
Motivated by these considerations, we  examine a mechanism by which excitons are formed  from 
an initial state of unbound electron-hole pairs to  account
for the observed short exciton formation time \cite{ceballos2016exciton} in
common TMDCs (MoS$_2$, MoSe$_2$, WS$_2$, and WSe$_2$).
While the focus of this paper is on the theoretical aspects of excitonic
interactions, the overall aim is to seek an understanding of the critical
factors that limit the exciton formation time which is of relevance
to experimental investigations involving device applications.

The unequal charges of the   basis atoms in   polar crystals allow 
a moving electron to polarize the electric field of its  surrounding medium.
The polarization effects  displaces the ions  giving rise to lattice vibrations of
 a optical phonon frequency  in resonance with the polarization field, and enable
direct Fr\"ohlich coupling between phonons  and  charge carriers. 
In this work we consider that the excitons are created   via the  two-dimensional 
Fr\"ohlich  interaction which provides a critical pathway by which 
charge carriers undergo energy loss to optical phonons  at elevated temperatures
in the monolayers MoS$_2$ and other transition-metal dichalcogenides \cite{kaasbjerg2014hot}.
The exciton is a neutral quasiparticle, and  polarization effects due to the 
longitudinal optical waves may appear to have less influence
than those associated with  polarization effects of the individual electron or hole.
 In reality  the internal state of the exciton  undergoes dipole type transitions and there occurs  measurable effects due to Fr\"ohlich  interactions in constrained systems.
The focus on LO phonons in the exciton formation process in this study is justified  by the large strength of excitonic interactions with high frequency phonons that arise due to the strong confinement of the exciton wave-functions in the real space of monolayer systems. Moreover the exciton-phonon
interaction is long ranged due to the existence of polarization effects  despite
 large separations between charge carriers and the ions in the material
system. The  phonon-limited  mobility is  largely dominated by polar optical scattering
via the Fr\"ohlich interaction  at room temperatures \cite{kaasbjerg12}.
Exciton formation may take place via deformation potential coupling
to acoustic phonons  \cite{oh2000excitonic,thilagam1993generation}, but is likely to occur
with less efficiency due to the  high exciton  binding energies
 \cite{makatom,chei12,ugeda2014giant,hill2015observation,chei12,komsa2012effects,thiljap}
in monolayer dichalcogenides.

In conventional semiconductors such as  the two band GaAs material system, excitons are formed  
via the Fr\"ohlich interaction  in the picosecond
time range   \cite{siantidis2001dynamics,oh2000exciton}. While excitons in
GaAs materials  are dominantly formed at the center of 
the Brillouin zone center,  the  formation process occurs at the non-central points in the momentum space
of monolayer TMDCs \cite{jones2013optical}. This gives rise to
quantitative differences in the exciton
 creation times between GaAs and TMDCs. For excitation energies higher than the band-gap of monolayer systems,
the electron-hole pair creates an exciton with 
a non-zero  wavevector  associated with its center-of-mass
motion \cite{siantidis2001dynamics,oh2000exciton}. The exciton subsequently  relaxes to the 
zero  wavevector state with  emission of acoustic or LO phonons
before  undergoing radiative recombination  by emitting
a photon. To this end, the formation time of an exciton as a function of exciton wave
vector is useful in  analyzing  the  luminescence rise times that can be
measured  experimentally.

In this study we employ the exciton-LO phonon
interaction operator to   estimate the exciton formation times in monolayer transition metal dichalcogenides.
The formation time of excitons is determined using
the interaction Hamiltonian which describes the 
conversion of the  photoexcited free electron-hole pair
to a final  exciton state initiated by exciton-phonon Fr\"ohlich interactions,
and  accompanied by absorption or emission 
of phonons.  The dependence of the 
 exciton formation time  on several parameters
such as the temperatures of the crystal lattice, charge carriers and
excitons as well as the densities of charge carriers and  excitons will be closely
examined in this study.
 
\section{Formation of Excitons in monolayer Molybdenum Disulfide \label{basic}}

\subsection{Exciton-LO phonon Hamiltonian}

 We project the single monolayer  of a hexagonally ordered plane of metal atoms sandwiched between two other hexagon planes of chalcogens onto a  quasi two-dimensional space \cite{cho2008,mouri2013}. The motion of the exciton is generally confined to the parallel two-dimensional $XY$  layers of the atomic planes with restricted electron and hole motion in the $z$ direction  perpendicular to the monolayer plane. 
The monolayer MoS$_2$  has nine
phonon branches consisting of three acoustic and six optical branches.
The two lowest optical branches  are weakly coupled to
the charge carriers are therefore not expected to play a significant role in the creation 
of excitons. The next two phonon branches  at the $\Sigma$ point positioned at energies 48 meV \cite{kaasbjerg12}
are linked to polar optical modes, which play a 
critical role in the formation of exciton after photoexcitation of the material system.
The  roles of the homopolar dispersionless mode at 50 meV  which typically occurs in layered
structures as well as the sixth phonon mode with the highest energy will  not be
considered here. Due to the large difference in momentum   between valleys in
TMDCs,  we assume that the exciton
formation occurs via an LO phonon-assisted intravalley process which preserves the valley
polarization in the monolayer system.

The Hamiltonian term associated with  the interaction between  excitons and 
 LO phonons is obtained by summing the electron-LO phonon and hole-LO phonon interaction
Hamiltonians as follows
\bea
H^{op}({\bf r_e,r_h}) &=&
{\sum_{\bf q}} {\mathcal C}
\left [  \exp (i {\bf q.r_e}) - \exp (i {\bf q.r_h}) b_{\vec{q},q_z}^{} + c.c \right ],\;
\label{hop} \\
 {\mathcal C} &=& \frac{i e}{|\vec{q}|} \sqrt{\frac{\hbar \omega_{LO}}{2 \epsilon_o V} (\frac{1}{\kappa_{\infty}}-
\frac{1}{\kappa_{0}})}\; {\rm erfc[} |\vec{q}| \; \sigma/2]
\label{coup}
\eea
where ${\bf{r_e}}=\left(x_e,y_e,z_e\right) = \left(\vec{r_e},z_e\right)$
and ${\bf{r_h}} = \left( x_h,y_h,z_h \right) = \left( \vec{r_h},z_h \right)$
denote the respective space coordinates of the electron and hole, and
$\vec{r_e}$ (or $\vec{r_h}$) marked with an arrow
denotes the monolayer in-plane coordinates of the electron (or hole). 
 The phonon creation and annihilation operators are denoted by
$b^{\dagger}_{\vec{q},q_z}$ and  $b_{\vec{q},q_z}$, respectively, where
 ${\bf{q}}=\left(\vec{q}, q_z\right)$  is composed
of the in-plane $\vec{q}$ and perpendicular  $q_z$ components of  the phonon wavevector.
The term $\omega_{LO}$ denotes the frequency of the LO phonon, $\epsilon_o$ is the permittivity of free space, $V$ is the volume of the crystal. The low-frequency and low-frequency relative dielectric constants are given by
$\kappa_{0}$ and $\kappa_{\infty}$,  respectively. The inclusion of the 
complementary error function $ {\rm erfc[q} \sigma/2]$ where $\sigma$ is the effective width of the electronic
Bloch states    is based on the constrained interaction
of  LO phonon with  charge carriers in  two-dimensional materials \cite{kaasbjerg12}.
For the monolayer MoS$_2$, the Fr\"ohlich coupling constant of  98 meV and an effective
width $\sigma$ = 4.41 \AA \; provide good  fit to the interaction energies 
evaluated from first principles  in the long-wavelength limit \cite{kaasbjerg12}.
 Due to dielectric screening,
the  Fr\"ohlich interaction decreases with increase in the phonon momentum,
and  larger coupling values $( \ge 330 )$ meV were obtained in the small momentum
limit in another study \cite{sohier2016two}. The Fr\"ohlich coupling constants obtained in
earlier works \cite{kaasbjerg12,sohier2016two}  will be used in this study to compute the
formation times of excitons.

The field operator $\hat\Psi^\dagger_{_{e-h}}$
of a pair of electron and hole with  a  centre of mass that moves freely
is  composed of electron and hole operators as follows
\be
\label{field}
 \hat\Psi^\dagger_{_{e-h}}({\vec{R}},{\vec{r}},z_e,z_h) 
= \frac{1}{A}  \sum_{{\vec{K}, \vec{k}}}
e^{- i\vec{K} \cdot \vec{R}} e^{- i\vec{k} \cdot \vec{r}} \psi_e(z_e) \; \psi_h(z_h) \;
a_{v,{\alpha_h \vec{K} - \vec{k}}}^{\dagger}\; a_{c,{\alpha_e \vec{K} + \vec{k}}}^{\dagger}\;,
\ee
where $A$ is the two-dimensional quantization area in the monolayer plane,
and  $a_{v,\vec{K}}^{\dagger}$ ($a_{c,\vec{K}}^{\dagger}$) are the respective hole
and electron creation operators with in-plane wavevector $\vec{K}$.
The  center-of-mass wavevector $\vec{K} = \vec{k_e}+\vec{k_h}$ and the wavevector of the
relative motion $\vec{k} = \alpha_h \vec{k_e}- \alpha_e \vec{k_h}$ where
 $\vec{k_e}$  ($\vec{k_h}$) is the electron (hole) in-plane wavevector,
with $\alpha_e ={m_e}/(m_e+m_h)$, $\alpha_h = {m_h}/(m_e+m_h)$
where $m_e$  ($m_h$ ) is the effective mass of the electron (hole).
In Eq.\ref{field}, the excitonic center of mass  coordinate $\vec{R}$ and relative
coordinate $\vec{r}$ parallel to the monolayer plane are given by
\bea
\label{coord}
\vec{R} &=& \alpha_e \vec{r_e} + \alpha_h \vec{r_h}\;, \\
\nonumber
\vec{r} &=& \vec{r_e} -\vec{r_h}.
\eea
The electron and hole wave functions ($\psi_e(z_e)$, $\psi_h(z_h)$) 
in the lowest-energy states are  given by $\mathcal{N} \; \cos[\frac{\pi z_j}{L_w}]$ (j = e,h)
for $|z_j| \le \frac{L_w}{2}$, and 0 for $|z_j| > \frac{L_w}{2}$. The term $\mathcal{N}$ denotes the normalization constant and 
$L_w$ is the average displacement of  electrons and holes in  the $z$ direction
perpendicular to the monolayer surface \cite{thilrelaxjap}.

\subsection{Exciton creation Hamiltonian}

The field operator $\hat\Psi^\dagger_{ex}$
of an exciton located at $\left(\vec{R},\vec{r},z_e,z_h \right)$
differs from the field operator $\hat\Psi^\dagger_{_{e-h}}$
of a free moving pair of electron and hole (see Eq.\ref{field}), and 
is given by \cite{taka2,oh2000exciton,thilagam2006spin}
\be
\label{fieldex}
 \hat\Psi^\dagger_{ex}({\vec{R}},{\vec{r}},z_e,z_h) 
= \frac{1}{A}  \sum_{\vec{K}}
e^{- i\vec{K} \cdot \vec{R}} \; \rho_{ex}^\star(\vec{r}) \; \psi_e(z_e) \; \psi_h(z_h) \;
B_{\vec{K}}^{\dagger},
\ee
where $B_{\vec{K}}^{\dagger}$ is the exciton creation operator with 
 center-of-mass wavevector $\vec{K}$ parallel to the monolayer plane.
The 1s two-dimensional exciton wavefunction  
$\rho_{ex}(\vec{r})$ = $\sqrt{\frac{2 \beta^2}{\pi}} \exp(- \beta |\vec{r}|)$ 
where $\beta$ is a variational parameter.
Using Eqs. \ref{hop}, \ref{field} and \ref{fieldex}, the Hamiltonian associated with the 
formation of an exciton from an initial state of free electron-hole
pair  with absorption/emission of   an LO phonon appear as
\bea
\label{int}
H^{F}_I &=& \frac{1}{\sqrt{A}}
\sum_{\vec{K},\vec{k},\vec{q},q_z} \; \lambda_o \; {\mathcal F_-}(\vec{k},\vec{q},q_z)\;
{\rm erfc[} \frac{|\vec{q}| \; \sigma}{2}]
B_{\vec{K}}^{\dagger}\; b_{\vec{q}, q_z} \;
a_{v,{\alpha_h (\vec{K}-\vec{q}) - \vec{k}}}\; a_{c,{\alpha_e (\vec{K}-\vec{q}) + \vec{k}}}\; \\
\nonumber
&+& \lambda_o \;  {\mathcal F_+}(\vec{k},\vec{q},q_z)\; {\rm erfc[} \frac{|\vec{q}| \; \sigma}{2}] \;
B_{\vec{K}}^{\dagger}\; b_{\vec{q}, q_z}^{\dagger} 
a_{v,{\alpha_h (\vec{K}+\vec{q}) - \vec{k}}}\; a_{c,{\alpha_e (\vec{K}+\vec{q}) + \vec{k}}}\;,
\\ \nonumber \\ \label{ffact}
{\mathcal F_{\mp}}(\vec{k},\vec{q},q_z) &=& {\mathcal F_e}(\pm q_z)\;  {\mathcal G}(\vec{K} \pm \alpha_h \vec{q})
- {\mathcal F_h}(\pm q_z) \; {\mathcal G} (\vec{K} \mp \alpha_e \vec{q}),
\\ \label{ffact2}
{\mathcal G}(\vec{K} \pm \alpha_i \vec{q}) &=& \int \; d\vec{r} \rho_{ex}^\star(\vec{r})\; 
e^{i(\vec{k} \pm \alpha_i \vec{q}) \cdot \vec{r}},
\\ \label{ffact3}
{\mathcal F_i}(q_z) &=& \int \; dz_i |\psi_i(z_i)|^2\; e^{i q_z z_j}, \; \; {\rm i = e,h}
\eea
where the coupling constant $\lambda_o = \sqrt{\frac{e^2 L_m \hbar \omega_{LO}}{2 \epsilon_o A} (\frac{1}{\kappa_{\infty}}- \frac{1}{\kappa_{0}})}$ and $L_m$ is the monolayer thickness. The form factor ${\mathcal G}$  is evaluated using the
explicit form of the two-dimensional exciton wavefunction  $\rho_{ex}(\vec{r})$.
Likewise  the second form factor ${\mathcal F}$ is computed using the
electron  wavefunction $\psi_e(z_e)$ and  hole wavefunction
 $\psi_h(z_h)$.

\subsection{Exciton formation rate}

For transitions involving a single phonon with wavevector $\vec{q}$,
the  formation rate of the exciton with wavevector $\vec{K}$
is computed by employing the Fermi golden rule
and the  interaction operator in Eq.\ref{int}
as follows
\bea
\label{rate}
W(\vec{K} \pm \vec{q},q_z) &=& \frac{1}{A} \frac{2 \pi}{\hbar} \;
|\lambda_o|^2 \; |{\mathcal F_{\pm}}(\vec{k},\vec{q},q_z)|^2 \; {\rm erfc[} \frac{|\vec{q}| \; \sigma}{2}]^2 \;
 f_h(\alpha_h (\vec{K} \pm \vec{q}) - \vec{k}) \;  f_e(\alpha_e (\vec{K} \pm \vec{q}) + \vec{k})
 \\
\nonumber &\times&  (f_{ex}(\vec{K})+1) \;
(\overline n_{\bf q} + \frac{1}{2} \pm \frac{1}{2}) \; \delta(E_{ex} - E_{eh}^\pm \pm \hbar \omega_{LO}),
\eea
where the emission (absorption) of phonon is denoted by $+$ ($-$), and the exciton energy 
$ E_{ex} =  E_g + E_{_b}+ \frac{\hbar^2 |\vec{K}|^2}{2 (m_e+m_h)}$  where $E_{_b}$ is the
exciton binding energy. The energies of charge carrier is  $E_{eh}^\pm = E_g + 
\frac{\hbar^2 |\vec{K} \pm \vec{q}|^2}{2 (m_e + m_h)} +  \frac{\hbar^2 |\vec{k}|^2}{2 \mu}$
where $\mu$, the reduced mass is obtained using $\frac{1}{\mu} = \frac{1}{m_e} + \frac{1}{m_h}$.
At low temperatures of the charge carriers  the phonon bath can be considered
to be thermal equilibrium with negligible phonon-phonon scatterings and phonon decay
processes.The  thermalized average occupation of phonons for low temperatures of the charge carriers is
given by 
\be
\label{phon}
{\overline n_{q}} = [\exp \left(\frac{\hbar \omega_{LO}}{k_B T_l}\right)-1]^{-1},
\ee
where  $T_l$ is the lattice
temperature and $\hbar \omega_{LO}$ is the energy of the LO phonon  that is emitted 
during the exciton generation process. 
The relaxation of electrons and holes  at high enough temperatures
($\approx 200 K$)  generally displaces   phonons beyond the equilibrium point when
 phonon-phonon related processes become dominant.
The phonon Boltzmann equation  \cite{kaasbjerg2014hot}
takes into account a common temperature that is achieved 
as a result of equilibrium reached between electrons and phonons. 
Hot-phonon effects is incorporated by  replacing
the temperature $T_l$ in Eq.\ref{phon} by an effective lattice temperature 
$T_{ph}$ \cite{kaasbjerg2014hot}.

The charge carriers are assumed to be in quasi-thermal equilibrium during the exciton formation
process. Consequently the occupation numbers ($f_h(K)), f_e(K)$) of hole and electron states in Eq.\ref{rate} 
can be  modeled using the Fermi-Dirac distribution 
\bea
\label{fd}
f^i(K_i) &=&\left[ \exp \left( \frac{E(K_i) - \mu_i}{K_B T_i}\right ) + 1 \right]^{-1}, \quad \; { i = e, h}\\
\label{fd2}
\mu_{i} &=& K_B T_i \ln \left[\exp \left(\frac{\pi \hbar^2 n_i}{m_i k_b T_i} -1 \right) \right],
\eea
where the  chemical potential $\mu_i$ is dependent on 
the  temperature $T_{i}$  and the two-dimensional density  $n_i$ of the charge carriers. 
When the mean inter-excitonic distance is higher
the exciton Bohr radius as considered to be the case in this study, 
the exciton can be assumed  to be an ideal boson \cite{thilpauli}
with a Bose-Einstein distribution \cite{ivanov1999bose}
\bea
\label{bes}
f^{ex}(K) &=&\left[ \exp \left( \frac{E(K) - \mu_{ex}}{K_B T_{ex}}\right ) - 1 \right]^{-1},\\
\label{bes2}
\mu_{ex} &=& K_B T_{ex} \ln \left[1-\exp \left(- \frac{2 \pi \hbar^2 n_{ex}}{g (m_e+m_h) k_b T_{ex}} \right) \right],
\eea
where $\mu_{ex}$ is the exciton chemical potential,  $T_{ex}$ is the exciton temperature 
and   $n_{ex}$ is the exciton density.  The  degeneracy factor $g$ is obtained 
as the product of the spin and valley degeneracy factors \cite{kaasbjerg12}.

\subsection{Numerical results of the Exciton formation Time}

The   formation time of an exciton with
wavevector $\vec{K}$, $T(\vec{K})$   is obtained  by summing the wavevectors $(\vec{k},\vec{q},q_z)$
over the rate obtained in Eq.\ref{rate}
\be
\label{formt}
\frac{1}{T(\vec{K})} = \sum_{\vec{k},\vec{q},q_z} \; W(\vec{K} \pm \vec{q},q_z)
\ee
To obtain quantitative estimates of  the exciton formation time using Eq.\ref{formt}, we use  the monolayer MoS$_2$ material parameters as $m_e$ = 0.51 $m_o$, $m_h$ = 0.58 $m_o$ \cite{jin2014intrinsic}
where $m_o$ is the free electron mass, and the  coupling constant $\alpha_o$=  330 meV \cite{sohier2016two}.
We set the phonon energy $\hbar \omega_{LO}$ = 48 meV \cite{kaasbjerg12}, and the layer thickness $h$ = 3.13 \AA \;  \cite{ding2011first} is used to determine
the  upper limit of $\approx$ 6 \AA \; for $L_w$, 
the average displacement of  electrons and holes in  the direction perpendicular to the plane of the
monolayer.  We fix the   effective lattice temperature 
$T_{ph}$ = 15 K, but vary the electron and  hole temperatures,
$T_e$ and $T_h$.

\begin{figure}[htp]
  \begin{center}
\subfigure{\label{Ka}\includegraphics[width=8.5cm]{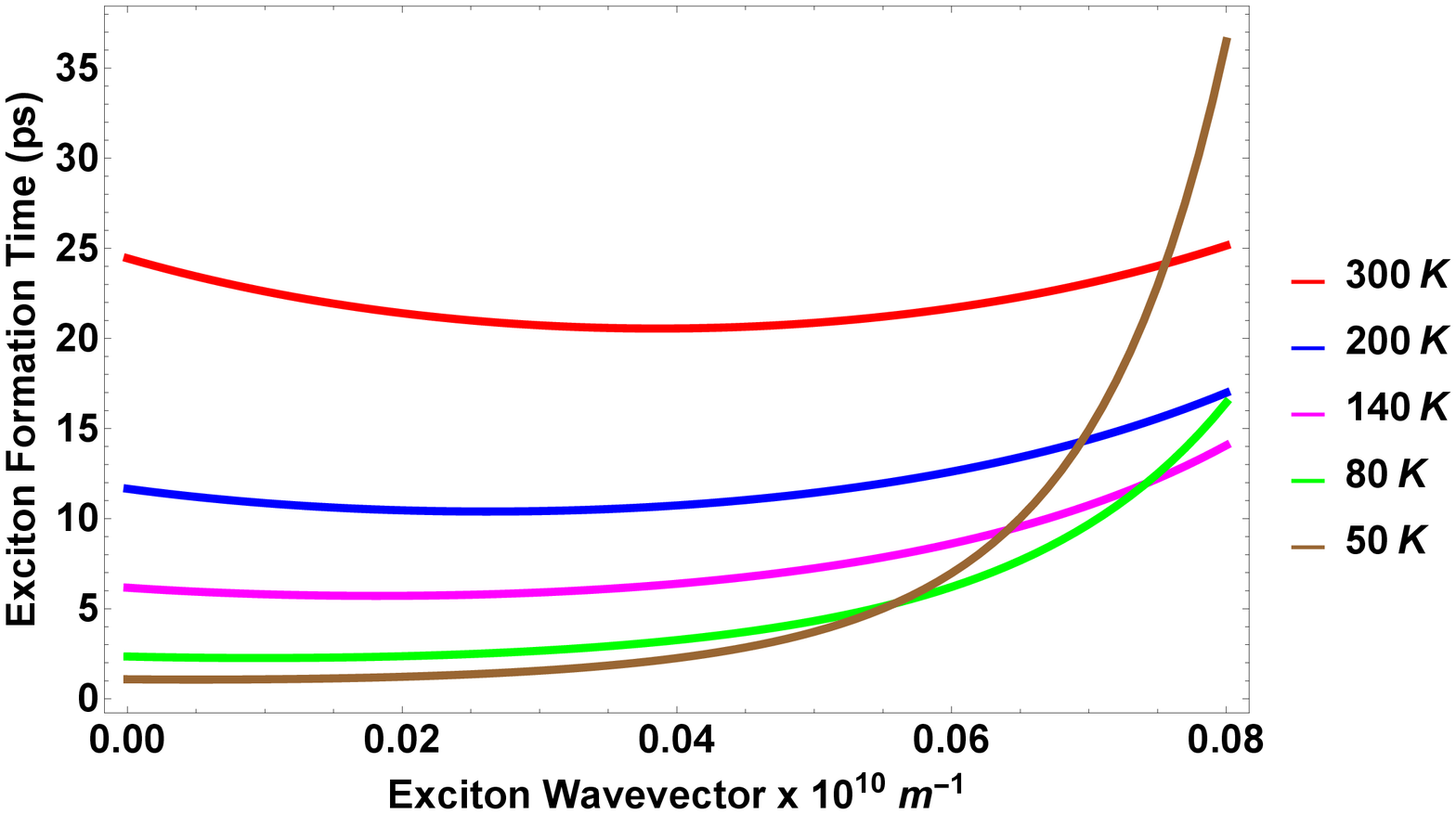}}\vspace{-1.1mm} \hspace{4.7mm}
\subfigure{\label{Kb}\includegraphics[width=8.5cm]{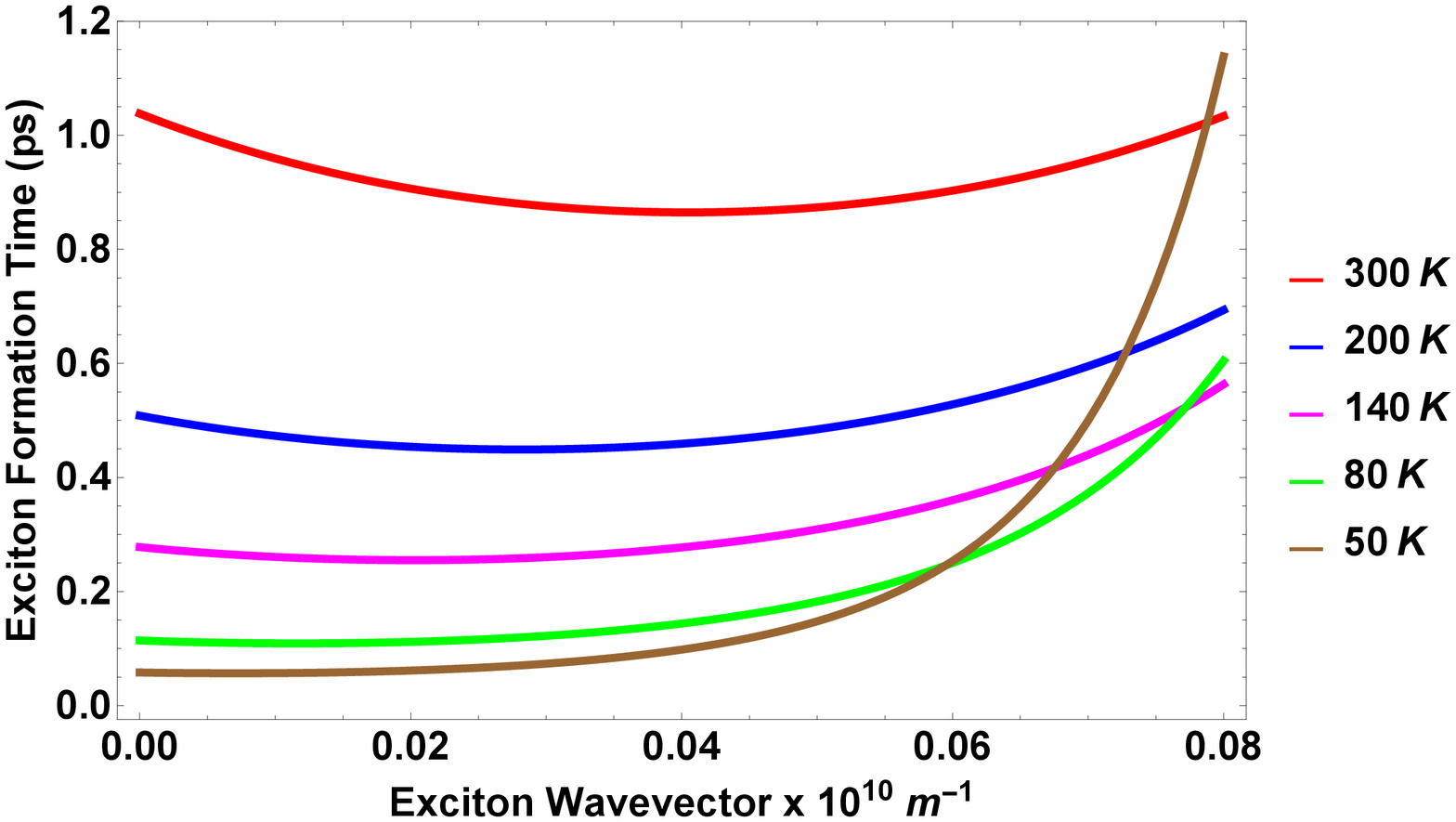}}\vspace{-1.1mm} \hspace{1.1mm} 
 \end{center}
     \caption{(a)
The exciton formation time as a function of the exciton center-of-mass wavevector $|\vec{K}|$
in the monolayer MoS$_2$ at different temperatures, $T_e$ = $T_h$ = $T_{ex}$ (50 K, 80 K, 140 K, 200 K, 300 K).
We use $m_e$ = 0.51 $m_o$, $m_h$ = 0.58 $m_o$ \cite{jin2014intrinsic}
where $m_o$ is the free electron mass, the  coupling constant $\alpha_o$ = 330 meV \cite{sohier2016two}
and $\hbar \omega_{LO}$ = 48 meV \cite{kaasbjerg12}.
The effective lattice temperature 
$T_{ph}$ = 15 K, $L_w$ $\approx$ 6 \AA \;, the exciton binding energy, 
$E_b$ = 300 meV \cite{thiljap} and densities, $n_e$ = $n_h$ = $n_{ex}$ = 1 $\times$ 10$^{11}$ cm$^{-2}$. \\
 (b) The exciton formation time as a function of the center-of-mass wavevector $|\vec{K}|$
in the monolayer MoS$_2$ at different temperatures, $T_e$ = $T_h$ = $T_{ex}$ (50 K, 80 K, 140 K, 200 K, 300 K).
All other parameters used are the same as specified in (a) with the exception of 
densities, $n_e$ = $n_h$ = $n_{ex}$ = 5 $\times$ 10$^{11}$ cm$^{-2}$.}
 \label{formK}
\end{figure}

Fig. \ref{formK}a,b show the calculated values (using Eqs. \ref{ffact}- \ref{formt})  of the exciton  formation
times as a function of exciton wavevector $\vec{K}$ with emission of an LO phonon
at different electron, hole and exciton  temperatures 
and densities, $n_e$ = $n_h$ = $n_{ex}$ = 1 $\times$ 10$^{11}$ cm$^{-2}$ and  5 $\times$ 10$^{11}$ cm$^{-2}$.
To obtain the results, we assume the temperatures to be the same for excitons and
unbound electron-hole pairs.
The results indicate that very fast 
 exciton formation times of less than one picosecond time occurs at
charge densities of 5 $\times$ 10$^{11}$ cm$^{-2}$ and carrier temperatures less than 300 K.
These  ultrafast sub-picosecond  exciton formation times  are in agreement with recent experimental findings  \cite{ceballos2016exciton} recorded at room temperatures in the monolayer MoS$_2$.
The exciton formation times are increased at the lower carrier densities of 1 $\times$ 10$^{11}$ cm$^{-2}$.

The wavevector of  exciton states  formed due to optical excitation 
of the ground state of the crystal  lies close 
to zero due to selection rules. The results in Fig. \ref{formK}a,b
show that while excitons are dominantly created at $|\vec{K}|$ = 0 at low charge carrier temperatures ($\approx$ 50 K), exciton formation occurs most rapidly at  non-zero exciton center-of-mass wavevectors 
($|\vec{K}|_f \neq$) at higher temperatures ($T_e$ = $T_h$ $\ge $ 140 K)  of the charge carriers.
At $T_e$ = $T_h$ $\approx $ 300 K, the shortest exciton formation time occurs at $|\vec{K}|_f$ = 
0.04 $\times$ 10$^{10}$ m$^{-1}$ (about 5.6 meV). 
The results in Fig. \ref{formK}a,b indicate that at  exciton wavevectors greater than 
0.06 $\times$ 10$^{10}$ m$^{-1}$, there is a notable increase in the exciton formation times linked to low electron-hole
plasma temperatures $T_e = T_h  \le$ 80 K. At high carrier temperatures there is likely
 conversion of newly formed composite bosons such as excitons into fermionic fragment species  \cite{thilagam2013crossover}.  The inclusion of considerations of
the quantum mechanical crossover of excitons into charge carriers at higher  plasma temperatures
will add to greater accuracy when computing exciton formation times.
This currently lies beyond the scope of this  study and will be considered in future
investigations where the role of the composite structure of excitons
in their formation rate will be examined.

%%%%%%%%%%%%%%%%%%%%%%%%%%%%%%%%%%%%%%%%%%%%%%%%%%%%%%%%%%%%%%%%%%%%%%%%%%%%%%%%%%%%%%%%%%%%

%%%%%%%%%%%%%%%%%%%%%%%%%%%%%%%%%%%%%%%%%%%%%%%%%%%%%%%%%%%%%%%%%%%%%%%%%%%%%%%%%%%%%%%%%%%%%%%%%%%%%%

\begin{figure}[htp]
  \begin{center}
\subfigure{\label{figa}\includegraphics[width=9.9 cm]{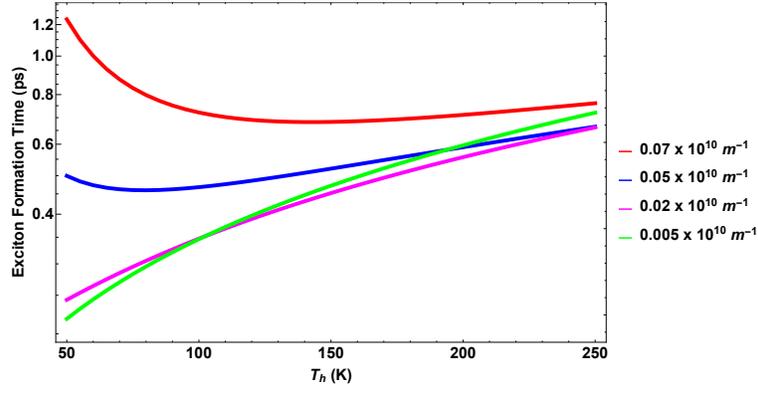}}\vspace{-1.1mm} \hspace{1.1mm} 
 \end{center}
     \caption{The exciton formation time as a  function of hole temperature
at different exciton center-of-mass wavevector $|\vec{K}|$ = 0.07 $\times$ 10$^{10}$ m$^{-1}$ (17.1 meV, red),
0.05 $\times$ 10$^{10}$ m$^{-1}$ (8.7 meV, blue), 0.02 $\times$ 10$^{10}$ m$^{-1}$ (1.4 meV, magenta), 
0.005 $\times$ 10$^{10}$ m$^{-1}$ (0.08 meV, green). The electron temperature is fixed at $T_e$ = 250 K,
exciton temperature $T_{ex}$ = 50 K and the carrier densities $n_e$ = $n_h$ = 5 $\times$ 10$^{11}$ cm$^{-2}$.
and $n_{ex}$ = 5 $\times$ 10$^{11}$ cm$^{-2}$. All other parameters used are the same as specified in the caption for Fig. \ref{formK}.}
 \label{diff}
\end{figure}

The effect of the variations within the electron-hole plasma temperatures or differences
between $T_e$ and $T_h$ on the exciton formation time is illustrated in Fig. \ref{diff}.
The formation times are computed at different exciton center-of-mass 
wavevectors  with the electron temperature fixed at $T_e$ = 250 K, and 
exciton temperature $T_{ex}$ = 50 K. The charge densities $n_e$ = $n_h$ = $n_{ex}$ = 5 $\times$ 10$^{11}$ cm$^{-2}$
and all other parameters used are the same as specified in the caption for Fig. \ref{formK}.
At the larger wavevector $|\vec{K}|$ = 0.07 $\times$ 10$^{10}$ m$^{-1}$ ($\approx$ 17.1 meV)
the formation time is shortest when the hole temperature  $T_h$ = $\approx $ 120.
With decrease in the center-of-mass wavevector $|\vec{K}|$, there is corresponding
decrease in the formation time when the hole temperature is allowed to decrease further from 
the electron temperature. At the  low exciton 
 wavevector $|\vec{K}|$ = 0.005 $\times$ 10$^{10}$ m$^{-1}$,
 the shortest formation time occurs  when the difference between $T_e$ and
$T_h$ reach the maximum possible value. These results  demonstrate the interplay of competitive effects of the hole-phonon and the electron-phonon dynamics on a picosecond time scale which
 results in a non-monotoinic temperature difference dependence $|T_h - T_e|$
of the exciton formation time.
%%%%%%%%%%%%%%%%%%%%%%%%%%%%%%%%%%%%%%%%%%%%%%%%%%%%%%%%%%%%%%%%%%%%%%%%%%%%%%%%%%%%%%%%%%%%

%%%%%%%%%%%%%%%%%%%%%%%%%%%%%%%%%%%%%%%%%%%%%%%%%%%%%%%%%%%%%%%%%%%%%%%%%%%%%%%%%%%%%%%%%%%%%%%%%%%%%%

\begin{figure}[htp]
  \begin{center}
\subfigure{\label{figa}\includegraphics[width=9.0cm]{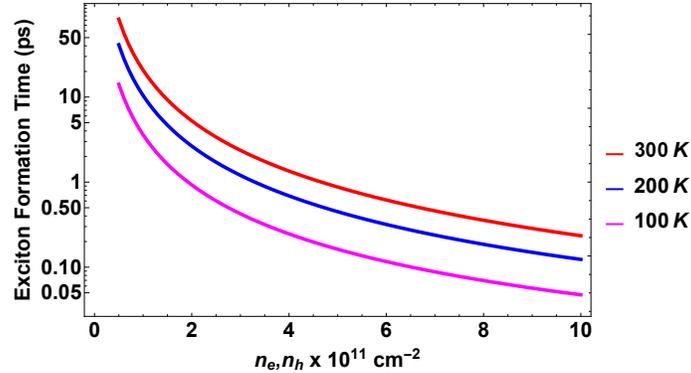}}\vspace{-1.1mm} \hspace{1.1mm} 
 \end{center}
     \caption{The exciton formation time as a function of the carrier density $n_e$ = $n_h$ at 
 different temperatures, $T_e$ = $T_h$ = $T_{ex}$ (300 K, 200 K, 100 K).
The exciton center-of-mass wavevector $|\vec{K}|$ = 0.03 $\times$ 10$^{10}$ m$^{-1}$.
All other parameters used are the same as specified in the caption for Fig. \ref{formK}.}
 \label{concf}
\end{figure}

In Fig. \ref{concf}, the exciton formation time is plotted 
as a function of the carrier density $n_e$ = $n_h$ at 
 different temperatures, $T_e$ = $T_h$ = $T_{ex}$ (300 K, 200 K, 100 K).
The exciton center-of-mass wavevector $|\vec{K}|$ = 0.03 $\times$ 10$^{10}$ m$^{-1}$.
All other parameters used are the same as specified in the caption for Fig. \ref{formK}.
Using the numerical values of the formation times, 
we performed numerical fits using the following relation which involve the carrier concentrations \cite{oh2000exciton}
\be
\label{fitT}
T(n_i) = \frac{B}{n_i^p} \quad \; { i = e, h}
\ee
where $B$ and $p$ are  fitting parameters. Using the results used
to obtain Fig. \ref{concf}, we get $B$ = 20.64 at $T_e$ = $T_h$ = 300 K,
$B$ = 10.35 at $T_e$ = $T_h$ = 200 K, $B$ = 3.56 at $T_e$ = $T_h$ = 100 K
and $B$ = 1.54 at $T_e$ = $T_h$ = 50 K. The constant $p \approx$ 2 irrespective
of the electron and hole temperatures. This implies an inverse 
square-law dependence of the exciton formation time 
 on the electron/hole concentration. Consequently a square-law
 dependence of the photoluminescence on excitation density is expected
to arise in the monolayer MoS$_2$ as well as other monolayer transition metal
dichalcogenides.

%%%%%%%%%%%%%%%%%%%%%%%%%%%%%%%%%%%%%%%%%%%%%%%%%%%%%%%%%%%%%%%%%%%%%%%%%%%%%%%%%%%%%%%%%%%%

%%%%%%%%%%%%%%%%%%%%%%%%%%%%%%%%%%%%%%%%%%%%%%%%%%%%%%%%%%%%%%%%%%%%%%%%%%%%%%%%%%%%%%%%%%%%%%%%%%%%%%

\section{Exciton formation times for other exemplary monolayer transition metal
dichalcogenides}

The theoretical results obtained in this study for MoS$_2$ are expected to be applicable to 
other low dimensional transition metal dichalcogenides. However subtle variations
 in the exciton formation times are  expected  due to differences in the exciton-LO coupling
strengths and energies of the LO phonon in the monolayer materials.
The  bare Fr\"ohlich interaction strengths obtained via ab initio techniques  
give 334 mev (MoS$_2$), 500 meV (MoSe$_2$), 140 mev (WS$_2$) and  276 meV (WSe$_2$) \cite{sohier2016two},
hence the Molybdenum-based TMDCs possess higher exciton-phonon coupling 
strengths than the Tungsten-based TMDCs. 
A precise estimate of the exciton binding energy in the 
 monolayer TMDCs  is not available, however a range of binding energies (100 to 800 meV) have been
reported for the monolayer systems \cite{hanbicki2015measurement,olsen2016simple,makatom,chei12,ugeda2014giant,
hill2015observation,choi2015linear,chei12,komsa2012effects,thiljap}. In order to compare
the exciton formation rates between Molybdenum-based TMDCs and 
Tungsten-based TMDCs, we make use of the effective masses of electron and holes 
at the $K$  energy valleys/peak position given in Ref. \cite{jin2014intrinsic}
and the Fr\"ohlich interaction strengths and 
 LO phonon energies  given in Ref.\cite{sohier2016two}. To simplify the numerical
analysis, we fix the exciton binding
energies at $\approx$ 330 meV for all the  TMDCs under investigation. This assumption is not expected
 to affect the order of magnitude of the exciton formation times, and also to
not detract from the analysis of effects of Fr\"ohlich interaction strengths
on the formation times.

\begin{figure}[htp]
  \begin{center}
\subfigure{\label{Ka}\includegraphics[width=8.5cm]{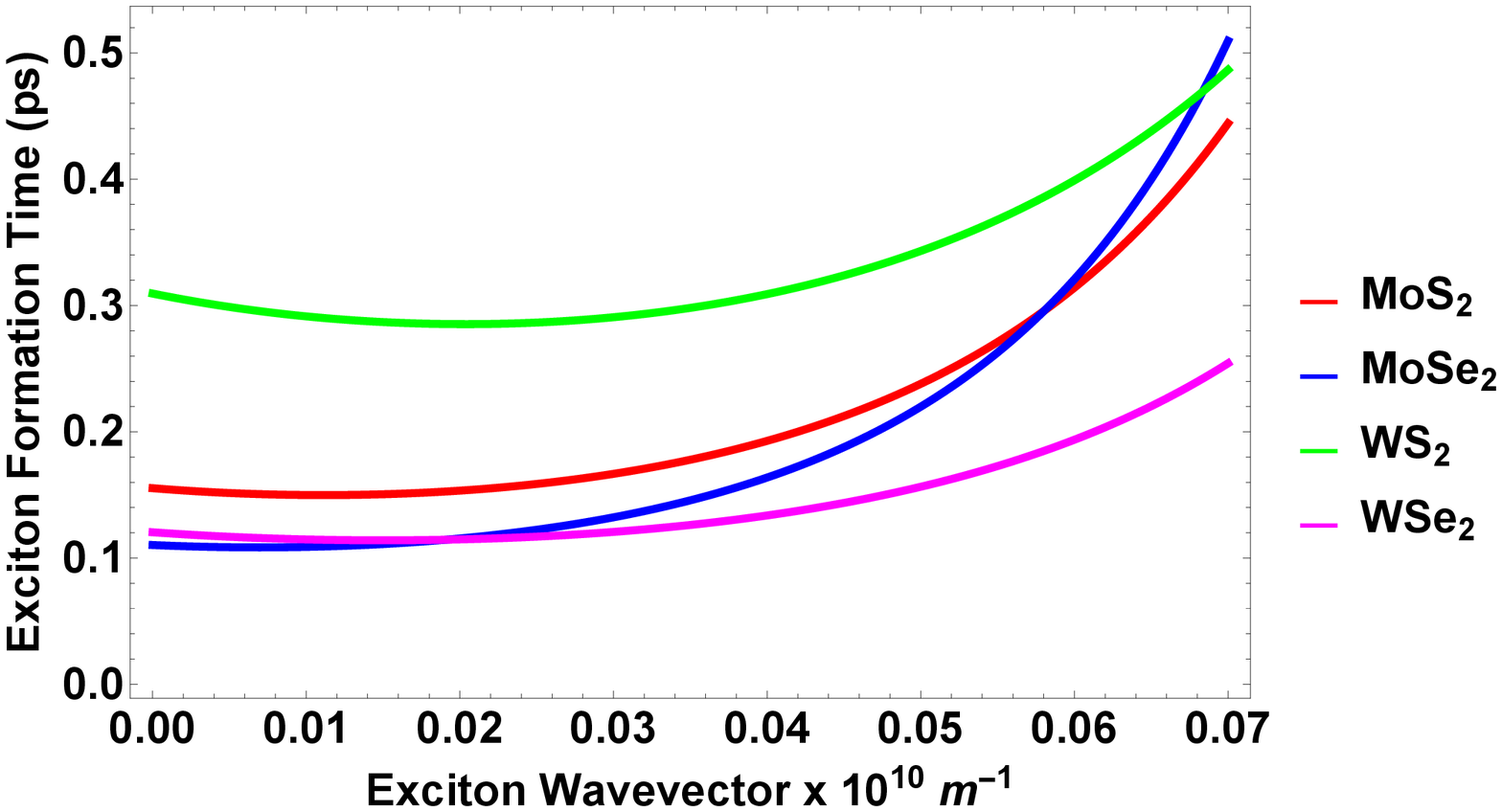}}\vspace{-1.1mm} \hspace{4.7mm}
\subfigure{\label{Kb}\includegraphics[width=8.5cm]{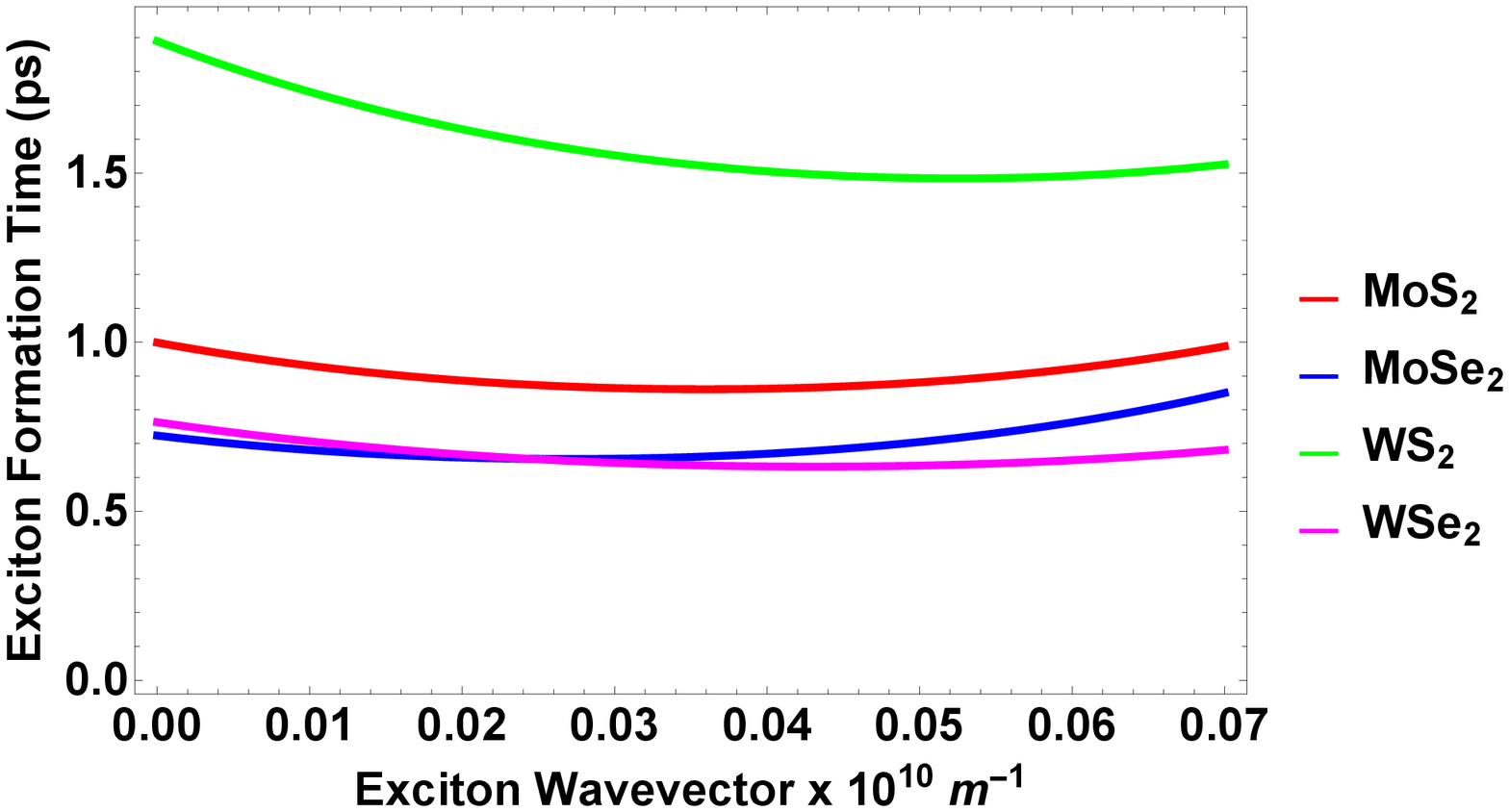}}\vspace{-1.1mm} \hspace{1.1mm} 
 \end{center}
     \caption{(a)
The exciton formation time as a function of the exciton center-of-mass wavevector $|\vec{K}|$
in  common monolayer systems (MoS$_2$, MoSe$_2$, WS$_2$, WSe$_2$)
 at  temperatures, $T_e$ = $T_h$ = $T_{ex}$ = 100 K. The effective masses of electron and holes 
at the $K$  energy valleys/peak are taken from Ref. \cite{jin2014intrinsic}
and the Fr\"ohlich interaction strengths and  LO phonon energies are obtained from Ref.\cite{sohier2016two}. 
The effective lattice temperature 
$T_{ph}$ = 15 K, $L_w$ $\approx$ 6 \AA \;, and 
densities, $n_e$ = $n_h$ = $n_{ex}$ = 5 $\times$ 10$^{11}$ cm$^{-2}$.\\
 (b) The exciton formation time as a function of the exciton center-of-mass wavevector $|\vec{K}|$
in common monolayer systems  at  temperatures, $T_e$ = $T_h$ = $T_{ex}$ = 300 K.
All other parameters used are the same as specified in (a) above.
}
 \label{comp}
\end{figure}

The results in Fig. \ref{comp} a, b show that the exciton formation times
of the selenide-based dichalcogenides 
are smaller than the  sulphide-based dichalcogenides at $T_e$ = $T_h$ = $T_{ex}$ = 100 K
and 300 K (with $|\vec{K}| \le$ 0.05 $\times$ 10$^{10}$ m$^{-1}$).
This is due to the comparatively higher Fr\"ohlich interaction strengths and lower LO phonon energies
of monolayer  MoSe$_2$ and WSe$_2$. The results in Fig. \ref{comp}a,b also indicate that
 excitons in the monolayer WS$_2$ are 
 dominantly created at non-zero  center-of-mass wavevectors  compared to the other three monolayer
dichalcogenide systems. This may be attributed to the comparatively lower effective masses
of hole and electron in the monolayer WS$_2$.

It is  instructive to compare
the exciton formation times in Fig. \ref{comp}a,b with the 
 radiative lifetimes of zero in-plane momentum 
excitons in suspended MoS2
monolayer of $\approx$ 0.18 - 0.30 ps  at 5 K 
\cite{wang2016radiative}.  The lifetimes of  excitons depend linearly on the
exciton temperature and increase  to the  picoseconds range at small temperatures and 
is larger than 1 ns at the room temperature. This indicates that the exciton formation processes
are likely to dominate  in the initial period when the TMDCs are optically excited
at high exciton temperatures. In the low temperature range (5 K - 20 K),
an interplay of competing effects of exciton generation and radiative decay
are  expected to occur  on the sub-picosecond time scale. 
Environmental parameters such as impurity concentration, exciton density and  density of 
excess charge carriers that affect the stability of  low dimensional trions will need
to be taken into account in order to  accurately model the exciton generation process at  the low temperature
regime.

The exciton formation scheme  adopted in this study has been parameterized by physical quantities such
as the exciton density and charge carrier densities. It is not immediately clear whether
these parameters can be extracted directly using ab-initio quantum mechanical and 
time-dependent density functional theory approaches.
Computations based on ab-initio techniques are generally numerically intensive and
time consuming  which are the main challenges in modeling low dimensional material systems. It is expected
that improvements in first principles modeling of anisotropic systems may result
in more efficient and rewarding approaches to determining the density functions of excitons
and charge carriers in future investigations. The  Auger process provides a non-radiative decay
channel for electron-hole pair recombination, hence this mechanism must be taken
into account for accurate  predictions of exciton formation times in future studies.

%%%%%%%%%%%%%%%%%%%%%%%%%%%%%%%%%%%%%%%%%%%%%%%%%%%%%%%%%%%%%%%%%%%%%%%%%%%%%%%%%%%%%%%%%%%%

%%%%%%%%%%%%%%%%%%%%%%%%%%%%%%%%%%%%%%%%%%%%%%%%%%%%%%%%%%%%%%%%%%%%%%%%%%%%%%%%%%%%%%%%%%%%%%%%%%%%%%

\section{Conclusion \label{conc}}

Transition metal chalcogenides have emerged as promising materials in which 
excitons exist as stable quasi-particles with high binding energies and thus play
important roles in the  optical processes of  monolayer TMDCs. 
The dynamics of excitons in monolayer transition metal
dichalcogenides  has been extensively studied  over the last
five years  in terms of both theory and applications. However
the  formation of excitons from free carriers has only been recently measured, and 
in this work we develop a model within the framework of Fermi's Golden rule to calculate the
formation dynamics of excitons from free carriers. 
This  theoretical study is aimed at providing 
a fundamental understanding of the exciton generation process
in optically excited monolayer transition metal dichalcogenides.
We focus on  a mechanism by which excitons are generated via 
the LO (longitudinal optical) phonon-assisted scattering process from  free electron-hole pairs 
in layered structures. The exciton formation time
is computed as a function of the  exciton center-of-mass wavevector,
electron and hole temperatures and  densities for known values of 
the  Fr\"ohlich coupling constant, LO phonon energy, lattice temperature  and  the exciton binding energy.
Our results  show that  the exciton is generated at
non-zero  wavevectors at higher temperatures ($\ge $ 120 K)  of  charge carriers,
that is  also  dependent on the density of the  electron and holes.
The inverse square-law dependence of the exciton formation time
on the density of charge carriers is also demonstrated by the
results of this study. 

For monolayer MoS$_2$, we obtain exciton formation times on the picosecond time scale at
charge densities of 1 $\times$ 10$^{11}$ cm$^{-2}$ and carrier temperatures less than 100 K.
The exciton formation times decreases to the sub-picosecond time range  at higher densities 
(5 $\times$ 10$^{11}$ cm$^{-2}$) and electron-hole plasma temperatures ($\le$ 300 K).
These ultrafast formation times are in agreement with recent experimental results
 ($\approx$ 0.3 ps) for  WSe$_2$, MoS$_2$, and MoSe$_2$ \cite{ceballos2016exciton}.
Due to the comparatively higher Fr\"ohlich interaction strengths and lower LO phonon energies
of monolayer  MoSe$_2$ and WSe$_2$, the exciton formation times
of the selenide-based dichalcogenides 
are smaller than the  sulphide-based dichalcogenides at $T_e$ = $T_h$ = $T_{ex}$ = 100 K
and 300 K (with $|\vec{K}| \le$ 0.05 $\times$ 10$^{10}$ m$^{-1}$).
The results of this study is expected to be  useful in understanding
the role of the exciton formation process in electroluminescence
studies \cite{sundaram2013electroluminescence,ye2014exciton} and exciton-mediated processes
in  photovoltaic  devices \cite{bernardi2013extraordinary,wi2014enhancement,tsuboi2015enhanced}.

\end{document}